\pdfoutput = 1
\documentclass[journal=nalefd, manuscript=letter]{achemso}
\usepackage{graphicx}
\usepackage{graphics}
\usepackage{amsmath}
\usepackage{amssymb}
\usepackage{amsfonts}
\usepackage{dcolumn}
\usepackage{dsfont}
\usepackage{latexsym}
\usepackage{rotating}
\usepackage{array}
\usepackage{makecell}
\usepackage{color}
\usepackage{latexsym}
\usepackage{bbm}
\usepackage{physics}
\usepackage{subfigure}
\usepackage{float}
\usepackage{epsfig}
\usepackage{epsf}
\usepackage{psfrag}
\usepackage{bm}
\usepackage{amsthm}
\usepackage{eucal}
\usepackage{mathrsfs}
\usepackage{tabularx}
\usepackage{url}
\usepackage{braket}
\usepackage{epstopdf}
\usepackage{color} 

\usepackage{hyperref}
\hypersetup{
colorlinks=true,
        final=true,
        linkcolor=black,
        citecolor=black,
        filecolor=black,
        urlcolor=black,
}

\title{Moir\'e skyrmions and chiral magnetic phases in twisted CrX$_{3}$ (X $=$ I, Br, Cl) bilayers}

\author{Muhammad Akram*}
\affiliation{Department of Physics, Arizona State University, Tempe, AZ 85287, USA}
\author{Harrison LaBollita*}
\affiliation{Department of Physics, Arizona State University, Tempe, AZ 85287, USA}
\author{Dibyendu Dey}
\affiliation{Department of Physics, Arizona State University, Tempe, AZ 85287, USA}
\author{Jesse Kapeghian}
\affiliation{Department of Physics, Arizona State University, Tempe, AZ 85287, USA}
\author{Onur Erten}
\affiliation{Department of Physics, Arizona State University, Tempe, AZ 85287, USA}
\email{onur.erten@asu.edu}
\author{Antia S. Botana}
\affiliation{Department of Physics, Arizona State University, Tempe, AZ 85287, USA}
\email{antia.botana@asu.edu}

\begin{document}


\begin{abstract}

We present a comprehensive theory of the magnetic phases in twisted bilayer Cr-trihalides through a combination of first-principles calculations and atomistic simulations. We show that the stacking-dependent interlayer exchange leads to an effective moir\'e field that is mostly ferromagnetic with antiferromagnetic patches. A wide range of noncollinear magnetic phases can be stabilized as a function of the twist angle and Dzyaloshinskii-Moriya interaction as a result of the competing interlayer antiferromagnetic coupling and the energy cost for forming domain walls. In particular, we demonstrate that for small twist angles various skyrmion crystal phases can be stabilized in both CrI$_3$ and CrBr$_3$. Our results provide an interpretation for the recent observation of noncollinear magnetic phases in twisted bilayer CrI$_3$ and demonstrate the possibility of engineering further nontrivial magnetic ground states in twisted bilayer Cr-trihalides.
\end{abstract}

Moir\'e superlattices arising from twisted bilayers of van der Waals (vdW) crystals represent an ideal platform for studying a plethora of novel phenomena: from the discovery of unconventional superconductivity in twisted bilayer graphene \cite{Cao2018} to the recent prediction of noncollinear magnetic states in two-dimensional (2D) magnetic materials \cite{Tong2018, Hejazi2020, Hejazi2021heterobilayer, akram2020skyrmions}. Among the currently known 2D vdW magnets, chromium trihalides (CrX$_{3}$, X = I, Br, Cl) represent a particularly interesting example in this context. These layered materials consist of ferromagnetic planes of Cr$^{3+}$: $d^3$ cations arranged in a honeycomb lattice with edge-sharing octahedral  coordination  \cite{Handy1952, Morosin1964, McGuire2015, Blei_APR2021}. 
Bulk CrX$_{3}$ materials exist in two structural phases with different stacking sequences along the $c$-axis: a low-temperature rhombohedral structure and a high-temperature monoclinic one \cite{McGuire2017, McGuire2015}. Recently, it has been shown that the stacking pattern in CrX$_3$ bilayers can 
give rise to a sign change in the interlayer magnetic coupling \cite{Sivadas2018, Klein2019, Gibertini_2020, Tong2021}. This stacking-dependent interlayer magnetic exchange has been experimentally demonstrated in both CrI$_3$ and CrBr$_3$, with the monoclinic stacking supporting an antiferromagnetic (AFM) interlayer coupling, while the rhombohedral stacking supports a ferromagnetic (FM) coupling instead \cite{Huang2017, Li2019, Song2019, Chen2019}. This result has important implications for the moir\'e physics:
a small twisting in bilayers creates a long-period moir\'e pattern in which the stacking order in each local region is similar to the corresponding lattice-matched stacking configuration. As such, the competing stacking-dependent interlayer interactions can give rise to novel magnetic states upon twisting, as has been recently demonstrated for CrI$_3$ \cite{Xu_arXiv2021}. 

\begin{figure}
    \centering
    \includegraphics[width=0.7\columnwidth]{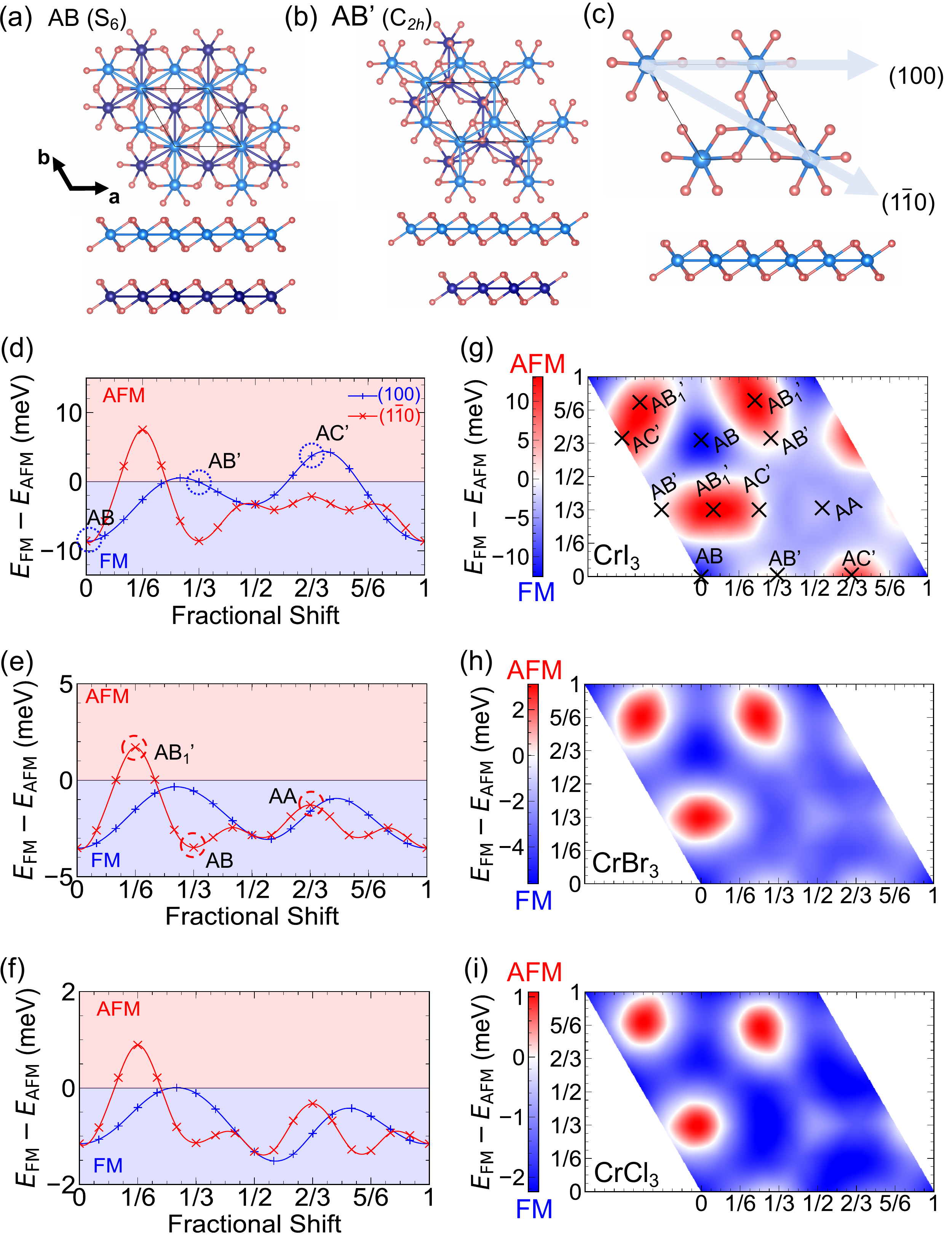}
    \caption{Top and side views of  CrX$_{3}$ bilayers are shown in panels (a) and (b) for a rhombohedral (AB) and monoclinic (AB') stacking, respectively. The unit cells are outlined in both cases. The larger, light (dark) blue spheres represent the Cr atoms in the top (bottom) layer while the smaller, red spheres represent the halide anions. (c) CrX$_{3}$ monolayer unit cell (top and side view) where the high-symmetry lateral shift directions [100] and [1$\bar{1}$0] are indicated by arrows. (d)-(f) Energy difference ($E_{\mathrm{FM}}-E_{\mathrm{AFM}}$) for various stacking displacements along the high-symmetry directions (100) (blue) and (1$\bar{1}$0) (red) for CrI$_{3}$, CrBr$_{3}$, and CrCl$_{3}$ bilayers, respectively. A positive energy difference denotes an AFM ground state while a negative difference denotes a FM ground state. (g)-(i) Moir\'e field patterns calculated for CrI$_{3}$, CrBr$_{3}$ and CrCl$_{3}$ bilayers, respectively. The patterns are obtained by interpolating the data in (d)-(f), via fitting to a polynomial. Blue regions correspond to a FM interlayer coupling, while red regions correspond to an AFM interlayer coupling. Specific stacking sequences defined in the main text are labelled in (g) and the corresponding lateral shifts are denoted in (d) and (e). The Supporting Information offers further details.}
    \label{fig:structures}
\end{figure}
In this letter, we study the magnetic phases in twisted bilayer CrX$_{3}$ spanning all three halide ions using a combination of {\it ab initio} calculations and atomistic simulations. We show that the interlayer exchange is mostly FM with three symmetry-related AFM patches. Noncollinear magnetic states, including skyrmion crystals, are obtained at small twist angles. These nontrivial phases arise  from the competition between the interlayer AFM coupling in the monoclinic stacking regions of the moir\'e superlattice and the energy cost to form AFM-FM domain walls. Our results agree with recent magnetic circular dichroism (MCD) experiments in CrI$_3$ \cite{Xu_arXiv2021} and provide insights for the wealth of noncollinear magnetic phases that can potentially be obtained in twisted bilayer Cr-trihalides.

We start by presenting the spin Hamiltonian that will enable the description of the magnetic properties of CrX$_{3}$ bilayers upon twisting:
\begin{equation}
\label{eqn:spinH}
\mathcal{H} = \mathcal{H}_{intra}^1+\mathcal{H}_{intra}^2+\mathcal{H}_{inter},
\end{equation}
where $\mathcal{H}_{intra}^{1 (2)}$ includes the symmetry-allowed intra-plane exchange terms in layer 1 (2) and $\mathcal{H}_{inter}$ incorporates the 
interlayer exchange, 
\begin{eqnarray} 
\label{Eq:Gnrl_Hmlt_1}
\mathcal{H}_{intra}&=& -\frac{J}{2}\sum_{{i,\mu}} {\bf S}_{i} \cdot {\bf S}_{i+\hat{\delta}_{\mu}}  -\frac{\lambda}{2}\sum_{{i,\mu}} {S}_{i}^{z} {S}_{i+\hat{\delta}_{\mu}}^{z} \nonumber\\
&&-\frac{D}{2} \sum_{{i,\mu}}[
\hat{d}_{\mu} \cdot ({\bf S}_{i} \times {\bf S}_{i+\hat{\delta}_{\mu}})]
-A_{s} \sum_{{i}}(S_{i}^{z})^2,\\
\label{eq:3}
\mathcal{H}_{inter}&=& -\sum_{\langle ij \rangle} J^\perp({\bf r}_{ij}) {\bf S}_i^{1}\cdot {\bf S}_j^{2}
\end{eqnarray}
here, $i$ is the site index and $\hat{\delta}_{\mu}$ are the three nearest-neighbors (nn) on the honeycomb lattice. $J$ is the intralayer
Heisenberg exchange coupling, $\lambda$  the anisotropic exchange coupling, and $A_s$ the single-ion anisotropy. $J^\perp({\bf r}_{ij})$ represents the  interlayer exchange coupling and ${\bf r}_{ij}$ is the interlayer displacement. 
Estimates for all these constants are obtained from first-principles calculations (see below). $D$ is the Dzyaloshinskii-Moriya  interaction (DMI) and is introduced in the atomistic simulations. 

The derivation of $J^\perp$ 
entails building the CrX$_3$ bilayers. Each bilayer is initially constructed using the  rhombohedral stacking with a vacuum of 20 \AA{} along the $c$-axis. The bilayers are subsequently relaxed within a FM state 
via  density functional theory (DFT) as implemented in the Vienna {\it ab-initio} Simulation Package (VASP) \cite{VASP} using projector augmented wave pseudopotentials \cite{PAW}. Different exchange-correlation functionals were attempted with both PBEsol \cite{PBEsol} and the DFT-D3 scheme \cite{DFTD3} (the latter including vdW interactions) 
giving an in-plane lattice parameter and interlayer distance that differ from experimental data less than 1\%. Based on this agreement, we proceed using the PBEsol functional throughout (see Supporting Information for further details).

For small twist angles, the stacking order at any local region in the moir\'e pattern can be obtained by translating one of the monolayer units in the rhombohedral reference cell by the vector $\vb{r} = \eta \vb{a} + \nu \vb{b}$, where $\eta, \nu\in [0, 1]$ and $\vb{a}, \vb{b}$ are the lattice vectors of the unit cell. We organize bilayer stacking sequences into two groups: symmetric stacking sequences 
(rhombohedral) denoted as AB, AA, and BA, and asymmetric stacking sequences 
(monoclinic) denoted as AB', AB$_{1}$', 
and AC'. All six of these stacking arrangements can be achieved from lateral shifts along two high-symmetry directions: (100) and (1$\bar{1}$0) (Fig. \ref{fig:structures}(a-c)). 
 
The moir\'e field is extracted from the stacking-dependent interlayer interaction in these bilayers \cite{Sivadas2018, Gibertini_2020, Tong2021}. $J^{\perp}({\bf r})$ in Eq. \ref{eq:3} for a given stacking is obtained as $(E_{\mathrm{AFM}}-E_{\mathrm{FM}})/2|{\bf S}|^2$, with $|{\bf S}|$= 3/2. Here, $E_{\mathrm{FM}}$ and $E_{\mathrm{AFM}}$ refer to the DFT energies in a FM state and an A-type AFM state. For the magnetic calculations at different stackings, we have employed DFT+$U$\cite{Liechtenstein1995} to treat the localized Cr-$3d$ electrons with an on-site Coulomb repulsion $U=3$ eV, consistent with previous literature,  \cite{Sivadas2018, Klein2019, Lu2020} and with cRPA calculations on CrI$_3$ \cite{mj} (see Supporting Information for further details on the DFT+$U$ calculations). The results for $E_{\mathrm{FM}}-E_{\mathrm{AFM}}$ at different stackings for CrX$_3$ bilayers are shown in  Fig. \ref{fig:structures}(d)-(i). As expected, the interlayer exchange modulates from FM to AFM as the stacking is changed -- a dominant interlayer FM coupling is obtained with three symmetry-related AFM patches. For all halides, the rotationally symmetric AB, AA, and BA stacking sequences strongly favor a FM interlayer coupling. However, the broken symmetry stackings (AB', AB$_{1}$', AC') differ between the three compounds. In CrI$_{3}$ an AFM coupling in all three stacking sequences is favored, while in CrBr$_{3}$ and CrCl$_{3}$ an AFM coupling is preferred only in the AB$_{1}$' stacking sequence. Accordingly, a systematic reduction in the size of the AFM regions can be observed in Fig. \ref{fig:structures} (g)-(i) as the size of the halide decreases. These findings agree with recent experimental \cite{Chen2019, Kim2019} and theoretical work \cite{Gibertini_2020, Sivadas2018, Soriano2019, Tong2021}. The sign change in the interlayer exchange can be understood from the competition between AFM nn $\mathrm{t}_{2g}-\mathrm{t}_{2g}$ couplings and FM next-nn $\mathrm{t}_{2g}-\mathrm{e}_{g}$ couplings. 
In the symmetric stacking sequences, the number of next-nn is greater than the number of nn leading to an overall FM exchange, while for the asymmetric stacking sequences the situation is reversed, leading to an AFM exchange \cite{Sivadas2018, Soriano2020}. First-principles-derived magnetic parameters in vdW magnets differ highly in the literature due to the use of different functionals and computational parameters \cite{Soriano2020}. We have employed a consistent methodology throughout to extract all couplings needed for the atomistic simulations (see Supporting Information for more details).  

To derive $J$, $\lambda$, and $A_{s}$ from first-principles,
we follow a procedure analogous to that employed in Ref. \citenum{Lado2017} for monolayer CrI$_3$. Specifically, we consider four magnetic configurations in CrX$_{3}$ monolayers: FM and AFM both oriented in-plane ($x$) and out-of-plane ($z$). These calculations are performed using the same methodology described above for the bilayers within PBEsol+$U$ ($U$= 3 eV) including spin-orbit coupling (SOC)  (see Supporting Information for the energy mappings). 
The derived values of $J$, $\lambda$, and $A_s$  are shown in Table \ref{tab:coefficients}. Importantly, we obtain $J>0$ (favoring FM interactions) in all materials with a value that decreases from I to Cl, as expected. 
The derived $A_s > 0$ favors an off-plane easy axis. For CrCl$_3$, this result disagrees with experiments \cite{McGuire2017}, so dipole-dipole interactions have to be included to turn its anisotropy in-plane (see Supporting Information for further details).

\begin{table}
\centering
\begin{tabular}{lccc}
\hline
\hline
 &  CrI$_3$ & CrBr$_{3}$ & CrCl$_{3}$\\
\hline
$J$ (meV)  & 4.06  & 3.42 & 2.37\\
$\lambda$ (meV)  & 0.14 & 0.04 & 0.003\\
$A_{s}$ (meV)  & 0.03  & 0.03 & 0.006\\

\hline
\hline
\end{tabular}
\caption{Magnetic parameters extracted from DFT calculations in monolayer CrX$_3$. $J$ is the intralayer symmetric Heisenberg exchange coupling, $\lambda$  the anisotropic exchange coupling and $A_s$ the single ion anisotropy. }
\label{tab:coefficients}
\end{table}

\begin{figure}
\centering
\includegraphics[width = 0.6\columnwidth]{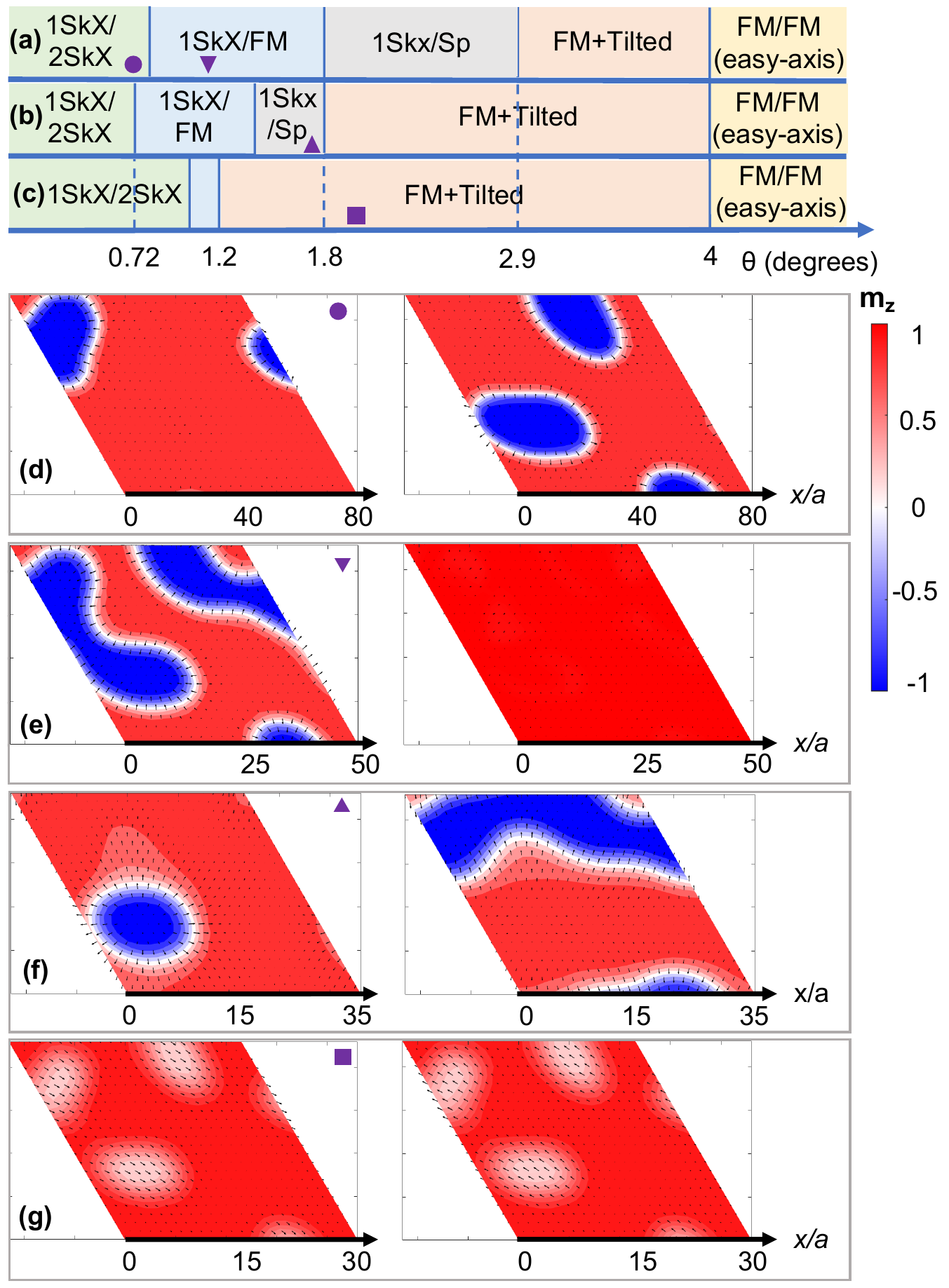}
\caption{Phase diagram and magnetization textures of twisted bilayer CrI$_3$. (a,b,c) Phase diagram of twisted bilayer CrI$_3$ as a function twisting angle $\theta$ for different values of DMI. Here (a), (b) and (c) correspond to $D/J=0.2$, $D/J=0.1$ and $D/J=0.05$ respectively. 
(d) Magnetization texture of 1SkX/2SkX for $D/J=0.2$ and $\theta=0.72^\circ$. The 1SkX/2SkX phase competes in energy with a 3SkX/FM state in which three skyrmions are formed on one layer while the other remains FM. The 1SkX/2SkX phase has slightly lower energy than the 3SkX/FM  since it minimizes the skyrmion-skyrmion interactions. Yet, this energy difference decreases for smaller $\theta$, as the distance between the skyrmions becomes larger. (e) Magnetization texture of 1SkX/FM for $D/J=0.2$ and $\theta=1.15^\circ$. (f) Magnetization texture of 1SkX/Sp for $D/J=0.1$ and $\theta=1.64^\circ$. (g) Magnetization texture of FM+Tilted state for $D/J=0.05$ and $\theta=1.91^\circ$. The x-axis in (d) to (g) shows the real space position within the moir\'e unit cell of size $L = a/ 2\sin(\theta/2)$ where a is the lattice spacing.}
\label{fig:CrI3}
\end{figure}

We now employ the magnetic parameters derived above and further introduce the DMI for the atomistic simulations. We introduce the DMI in this manner as its first-principles derivation would entail non-collinear calculations beyond the monolayer (where DMI is not allowed by symmetry). Also, this allows us to easily probe changes upon tuning the DMI, something that could be attainable experimentally  by utilizing different substrates or via liquid ion gating\cite{Diez_PRA2019}. As such, we consider three different values of the DMI for each material: $D/J=0.05, 0.1$ and $0.2$. 
In order to find the ground state for the bilayer Hamiltonian (Eq. (1)), we take the continuum limit and solve the Landau-Lifshitz-Gilbert (LLG) equations: \cite{1353448}
\begin{eqnarray} 
\frac{d\textbf{m}}{dt}=-\gamma \textbf{m}\times \textbf{B}^{eff}+\alpha \textbf{m} \times \frac{d\textbf{m}}{dt},
\end{eqnarray}
where  $\textbf{m}$ is the magnetization, $\textbf{B}^{eff}=-\delta \mathcal{H} /\delta \textbf{m}$, $\gamma$ is the gyromagnetic ratio and $\alpha$ is Gilbert damping coefficient. We solve the LLG equations for each layer self-consistently keeping 
$|{\bf m}|=1$
and imposing periodic boundary conditions.  Our method is suitable to capture magnetic phases that are commensurate with the moir\'e superlattice \cite{Hejazi2021heterobilayer} (See Supporting Information for further details). 

\begin{figure}
\centering
\includegraphics[width = 0.5\columnwidth]{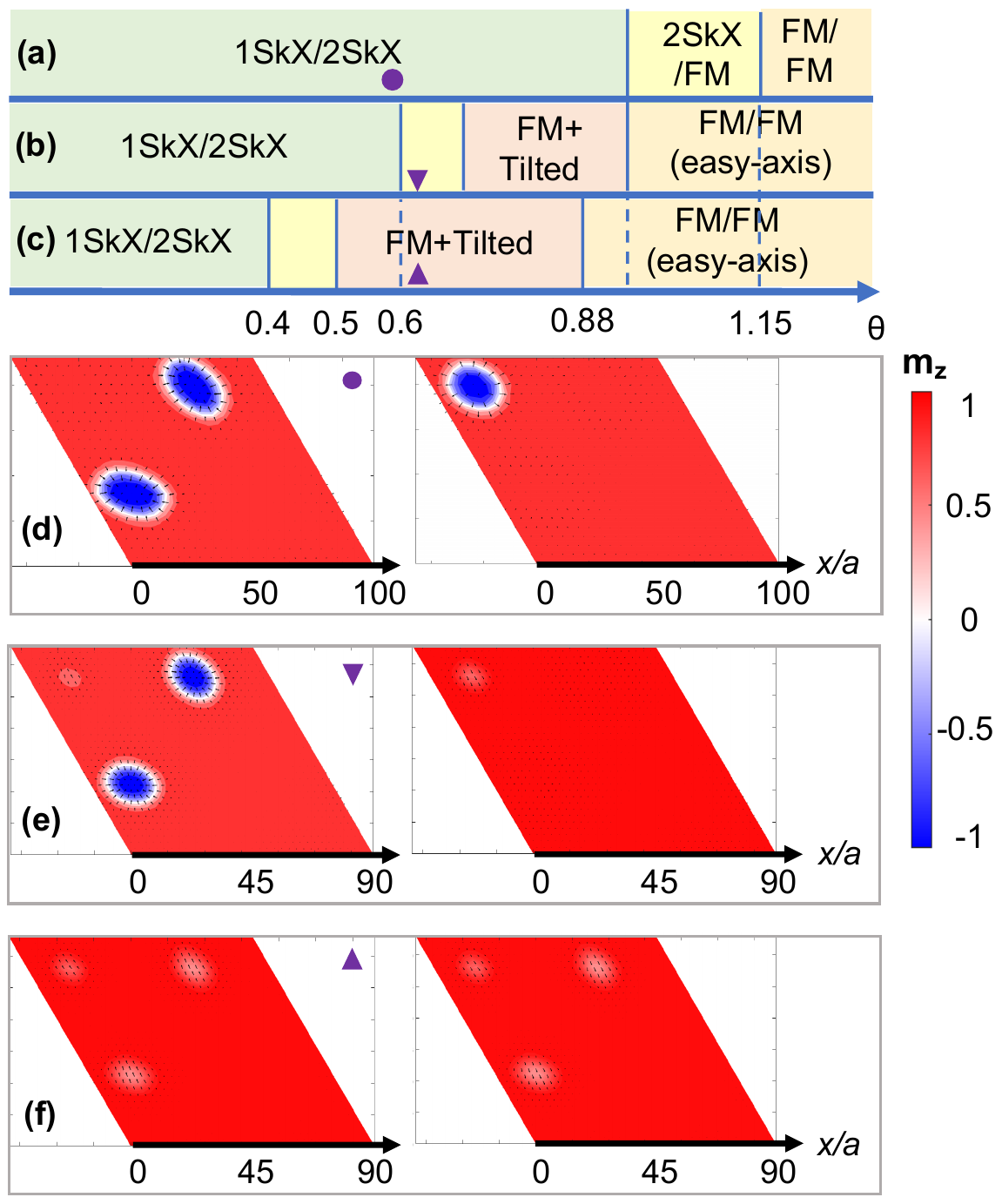}
\caption{Phase diagram and magnetization textures in twisted bilayer CrBr$_3$. (a,b,c) Phase diagram of twisted bilayer CrBr$_3$ as a function of twist angle $\theta$ for different values of DMI. Here (a),(b) and (c) correspond to $D/J=0.2$, $D/J=0.1$ and $D/J=0.05$ respectively. (d) Magnetization texture of 1SkX/2SkX for $D/J=0.2$ and $\theta=0.57^\circ$. (e) Magnetization texture of 2SkX/FM for $D/J=0.1$ and $\theta=0.64^\circ$. (f) Magnetization texture of the FM+Tilted state for $D/J=0.05$ and $\theta=0.64^\circ$. The x-axis in (d) to (f) shows the real space position within the moir\'e unit cell.}
\label{fig:CrBr3}
\end{figure}

\begin{figure}
\centering
\includegraphics[width = 0.5\columnwidth]{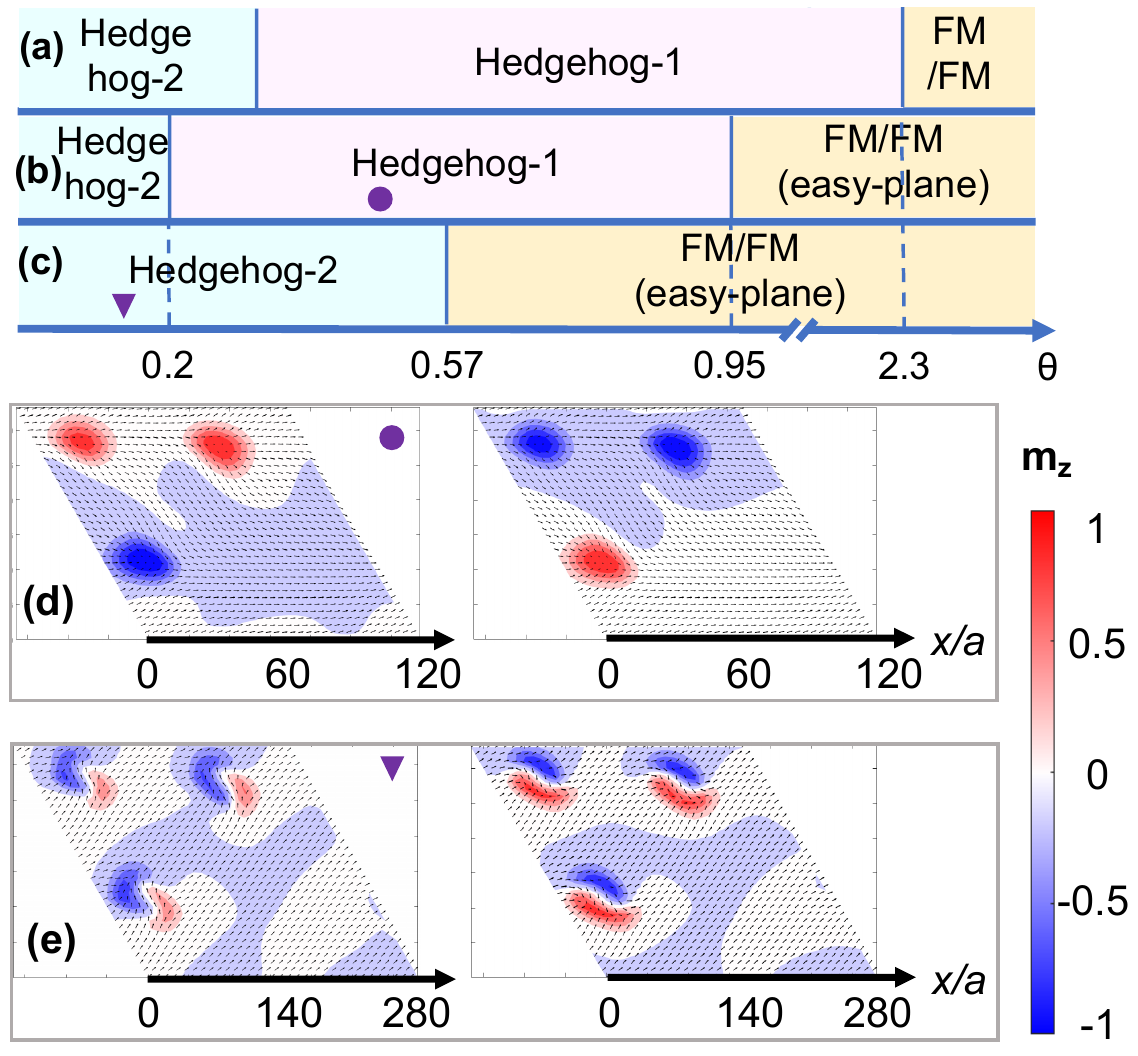}
\caption{Phase diagram and magnetization textures of twisted bilayer CrCl$_3$. (a,b,c) Phase diagram of twisted bilayer CrCl$_3$ as a function of twist angle $\theta$ for different values of the DMI. Here (a),(b) and (c) correspond to $D/J=0.2$, $D/J=0.1$ and $D/J=0.05$ respectively. (d) Magnetization texture of Hedgehog-1 for $D/J=0.1$ and $\theta=0.48^\circ$. (e) Magnetization texture of Hedgehog-2 for $D/J=0.05$ and $\theta=0.2^\circ$. The x-axis in (d) and (e) shows the real space position within the moir\'e unit cell.}
\label{fig:CrCl3}
\end{figure}

The main results for the magnetic phase diagrams at T=0 are summarized in Figs. \ref{fig:CrI3}, \ref{fig:CrBr3}, and \ref{fig:CrCl3} for CrI$_3$, CrBr$_3$, and CrCl$_3$, respectively. A plethora of noncollinear phases, including skyrmion crystals (SkX), are obtained at small angles (or large moir\'e periods) driven by the competing magnetic interactions in the moir\'e superlattices. 
Ref. \citenum{Tong2021} has also studied the magnetic phase diagrams of twisted trihalide bilayers 
but the DMI was not considered, impeding the stabilization of SkX phases. 

We start our discussion of the atomistic simulations with CrI$_3$ bilayers that display the richest phase diagram as a function of the twist angle (Fig. \ref{fig:CrI3}(a)-(c)). For the smallest angles, 
a 1SkX/2SkX phase is the ground state
(Fig. \ref{fig:CrI3}(d)). This phase is stabilized when the moir\'e superlattice is large enough so that one skyrmion forms in an AFM region on one layer while two skyrmions form in the remaining two AFM patches on the other layer. For $\theta$ $>$ 0.72$^\circ$, a 1SkX/FM phase is obtained with a single skyrmion forming in the three overlapping  AFM regions  on  one layer  while the  other stays  FM (Fig. \ref{fig:CrI3}(e)). For even larger angles, a noncollinear 1 SkX/spiral (Sp) phase is formed with a single skyrmion in one of the AFM patches on one layer and a spiral in the remaining two AFM regions on the other layer (Fig. \ref{fig:CrI3}(f)). As  the angle is increased further, a FM+Tilted phase is obtained in which the magnetization is out-of-plane in the FM background but it acquires a finite in-plane component in the AFM regions (Fig. \ref{fig:CrI3}(g)). The in-plane magnetization is opposite in the two layers and its magnitude increases with decreasing angle. Finally, a transition from the FM+Tilted phase to a FM/FM state is obtained at $\theta\sim4^\circ$. This FM/FM phase is the ground state at large angles for all  DMI strengths. 

An analytical estimate for the critical twist angle to obtain a skyrmionic phase transition ($\theta_{c}$) can be derived considering the competition between domain wall energy formation and interlayer exchange energy.  
Considering a single AFM patch, flipping the magnetization in one of the layers would lower the interlayer exchange energy producing an energy gain $E_{AFM} \sim 2 f_{AFM}\left(\frac{L}{a}\right)^2 \bar{J}^{\perp}$ ($f_{AFM}$ is the area fraction of the AFM patch over the moir\'e unit cell, $\bar{J}^{\perp}$ is the average AFM coupling and $a$ is the lattice spacing). This energy scales with the area of the AFM patch and therefore varies quadratically with the moir\'e period, $L$. On the other hand, this creates a domain wall with the rest of the system of length $\delta \sim \pi \big((J+\frac{\lambda}{2})/(A_s+\lambda+0.72 \bar{J}^{\perp}) \big)^{1/2}a$ and energy cost $E_{DW} \sim \left(\frac{\pi }{a} \sqrt{\left(J+\frac{\lambda}{2}\right)\left(A_s+\lambda+0.72 \bar{J}^{\perp}\right)}-\frac{D\pi }{a}\right)L$. Similar expressions are obtained in Ref. \citenum{Xu_arXiv2021} but we also consider the contribution from the DMI (see Supporting Information for more details). Importantly, for large enough $L$, $E_{AFM}$ can overcome $E_{DW}$ and lead to a phase transition to a skyrmionic phase. The critical angle for the transition is determined by minimizing the energy.
\begin{eqnarray}
\theta_{c} \sim \frac{2f_{AF} \bar{J}^{\perp}}{\pi \left( \sqrt{\left(J+\frac{\lambda}{2}\right)\left(A_s+\lambda+0.72 \bar{J}^{\perp}\right)}-D\right)}
\label{eq:theta_c}
\end{eqnarray}

This estimate is for a single domain and therefore only applicable when the AFM patches are well separated. When the domain wall length $\delta$ is comparable to the separation between AFM patches (as in CrI$_3$, Fig. \ref{fig:structures}(g)), there can be more complicated states. 
In any case, the analytical estimate in CrI$_3$ ($\theta_c$  $\sim3^\circ$ ($D/J=0.2$) and $2.3^\circ$ ($D/J=0.1$)) is in good agreement with our numerical results for the 1SkX/Sp to FM+Tilted phase boundary ($\theta_c \sim2.9^\circ$ ($D/J=0.2$) and $1.8^\circ$ ($D/J=0.1$)). The trends are the expected ones as an increasing DMI decreases the domain wall energy and therefore increases $\theta_c$.

\begin{figure}
\centering
\includegraphics[width = \columnwidth]{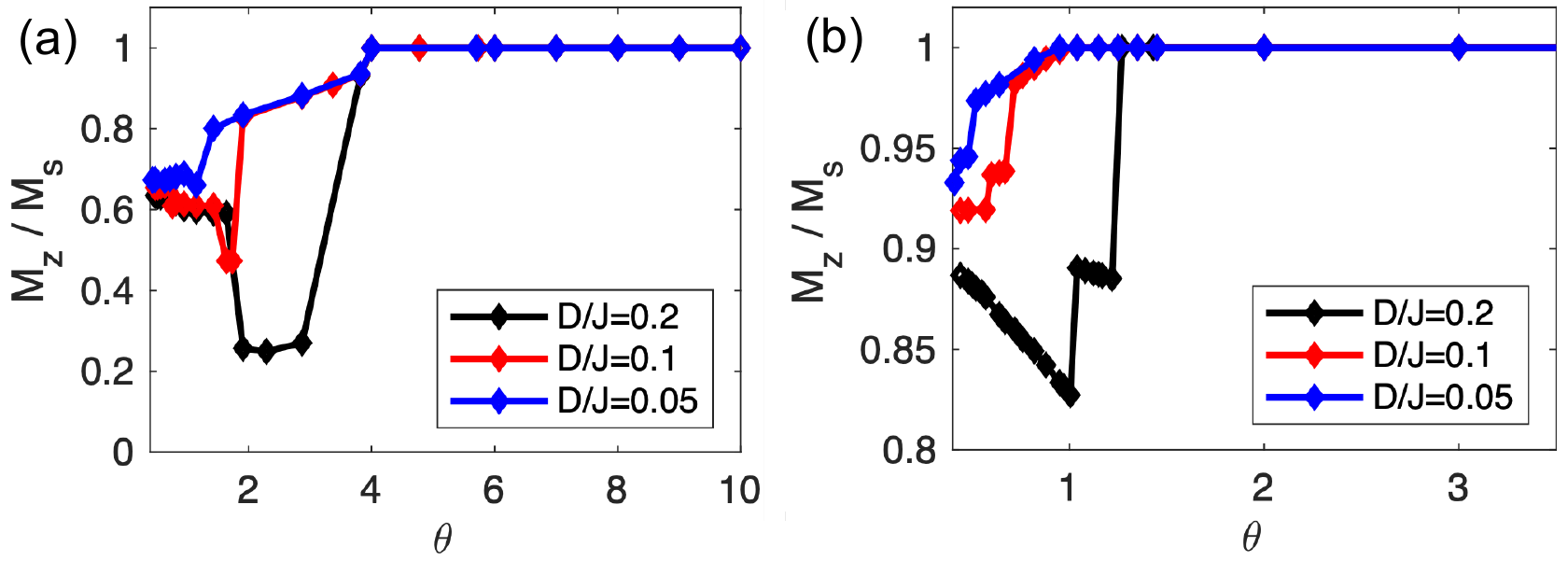}
\caption{Z-component of the normalized average magnetization as a function of twist angle ($\theta$) for (a) CrI$_3$, and (b) CrBr$_3$ for varying $D/J$.}
\label{fig:Mztheta}
\end{figure}

Our phase diagram is in agreement with recent experiments on CrI$_3$ bilayers on hexagonal BN (h-BN) substrates that demonstrated their high degree of magnetic tunability upon twisting, gating, and applied magnetic field using MCD \cite{Xu_arXiv2021}. A FM/FM phase is indeed the ground state in these experiments for $\theta > 3^\circ$, while a noncollinear magnetic ground state arises at smaller angles, in agreement with our results. Since these MCD experiments are only sensitive to the z-component of the magnetization, further work will be necessary to pinpoint the nature of this noncollinear magnetic phase and to check for the skyrmion phases we predict.

In order to make further connections with Ref. \citenum{Xu_arXiv2021}, we have also calculated the ratio of the z-component of the normalized magnetization ($M_z$) and the saturation magnetization ($M_s$) for bilayer CrI$_3$ (Fig.~\ref{fig:Mztheta}(a)). Experiments show that below $\theta \sim 3^\circ$ $M_z/M_s$ drops gradually to a saturation value of $\sim 0.6$ \cite{Xu_arXiv2021}. Our simulations show a transition of the same nature taking place for $\theta \sim4^\circ$ with $M_z$ slowly decreasing into the FM+Tilted phase as the angle is lowered. For our lowest and intermediate DMIs ($D/J=0.05$ and 0.1), the magnetization 
decreases gradually into the different skyrmion phases, until it saturates at small angles to $M_z/M_s \sim0.65$, close to the experimental value. For the largest DMI ($D/J=0.2$), the magnetization drops abruptly in the 1SkX/Sp phase instead, recovering the  $M_z/M_s \sim0.65$ value at low angles. This abrupt drop has not been observed in experiments \cite{Xu_arXiv2021}, suggesting the DMI is smaller in bilayer CrI$_3$ on h-BN substrates. 

Next, we discuss the phase diagram of CrBr$_3$. The ground states are similar to those in CrI$_3$ with FM/FM and 1SkX/2SkX phases emerging at large and small angles, respectively  (Fig.~\ref{fig:CrBr3}(a)-(d)). At intermediate angles, 2SkX/FM (Fig.~\ref{fig:CrBr3}(e)) and FM+Tilted (Fig.~\ref{fig:CrBr3}(f)) states are stabilized. The 2SkX/FM phase has two skyrmions on one layer along with a tilted-FM on the third AFM patch. 
CrBr$_3$ satisfies the assumptions for the analytical estimate of the critical angle using Eq.~\ref{eq:theta_c} better than CrI$_3$ as the size of the AFM patches is smaller and they are well separated (Fig. \ref{fig:structures}(h)). This estimate gives $\theta_c$ for the skyrmionic phase transition $\sim0.5^\circ$, $0.7^\circ$, and $1.4^\circ$  for $D/J=0.05$, 0.1 and 0.2, respectively, consistent with the 2SkX/FM to FM/FM phase boundary that takes place at $\theta_c\sim0.5^\circ$, $0.7^\circ$, and $1.15^\circ$. Given the good agreement achieved between our $M_z/M_s$ calculations and experimental data in CrI$_3$, we also calculate $M_z/M_s$ for CrBr$_3$. The overall magnetization drop in this case is much less pronounced (different scale between left and right panels in Fig. \ref{fig:Mztheta}) as the AFM patches are smaller than in CrI$_3$. The overall angle dependence remains similar for the different $D/J$ values, with the magnetization dropping gradually for $D/J$= 0.05 and 0.1, while a more abrupt change is obtained for $D/J$= 0.2. At small angles, saturation is reached at $M_z/M_s \sim0.9-0.95$.  Based on our results, twisted  bilayer CrBr$_3$ is a promising system to study as it should also display a rich phase diagram with nontrivial magnetic phases.

Finally, we discuss the phase diagram for CrCl$_3$ (Fig.~\ref{fig:CrCl3}(a)-(c)). Similar to CrI$_3$ and CrBr$_3$, a FM/FM phase is the ground state for large angles with the distinction that in this case the magnetization is in-plane
which prevents SkX formation at small angles.
Instead, two noncollinear magnetic orders arise, labelled as Hedgehog-1 and Hedgehog-2 (Fig.~\ref{fig:CrCl3}(d)-(e)), in which the in-plane magnetization acquires an out-of-plane component around the AFM patches that is opposite in the two layers. Hedgehog-1 resembles a meron configuration as the magnetization points up (or down) at the cores, and in-plane at the edges. However, unlike merons (or antimerons), the magnetization does not wind at the edges and therefore it does not have a half-integer quantized topological charge. The Hedgehog-2 configuration has similar features with a single AFM patch resembling a meron-antimeron pair. However, it also does not have the proper in-plane winding of the spins and therefore it does not have a quantized topological charge. The topological charge distribution of these two phases is presented in the Supporting Information. 

In conclusion,  we have shown that the interplay of the stacking-dependent interlayer exchange, twist angle, and DMI can lead to noncollinear magnetic states in bilayer Cr-trihalides. In particular, we demonstrate that for small twist angles various skyrmion crystal phases can be stabilized in both CrI$_3$ and CrBr$_3$, whereas for large angles all three systems are ferromagnetic. While we have focused here on homobilayer Cr-trihalides, our approach provides a general framework for understanding chiral magnetism in twisted 2D magnets. Interesting directions for future work include the effects of gating and external magnetic field, as well as the study of heterobilayers (CrX$_3$/CrX$^\prime_3$) combining different halide ions. In this situation, the lattice mismatch will give rise to a moir\'e pattern even without twisting. The effects of strain\cite{strain1, strain2} and interlayer separation\cite{Klein2019} are also interesting paths to explore.

\section{Associated Content}

\noindent {\bf Supporting Information}

\noindent Structural parameters, comparison of exchange-correlation functionals, magnetic shape anisotropy, continuum free energy, topological charge density of CrCl$_3$, domain wall energy and critical angle estimation (pdf).

\section{Author Information}
\noindent {\bf Corresponding Author:}
Onur Erten -- Email: onur.erten@asu.edu

\noindent {\bf Author Contributions:}
* Muhammad Akram and Harrison LaBollita contributed equally.

\noindent {\bf Notes:} The authors declare no competing financial interest.

\section{Acknowledgements}
We thank Nikhil Sivadas for sharing the data published in Ref.~\citenum{Sivadas2018} with us. JK, OE and AB acknowledge support from National Science Foundation Award No. DMR 1904716. MA is supported by Fulbright Scholarship. We acknowledge the ASU Research Computing Center for HPC resources. 

\bibliography{ref_resub.bib}
\end{document}


\title{Supporting Information for ``Moir\'e skyrmions and chiral magnetic phases in twisted CrX$_{3}$ (X = I, Br, Cl) bilayers''}
\author{Muhammad Akram*}
\affiliation{Department of Physics, Arizona State University, Tempe, AZ 85287, USA}
\author{Harrison LaBollita*}
\affiliation{Department of Physics, Arizona State University, Tempe, AZ 85287, USA}
\author{Dibyendu Dey}
\affiliation{Department of Physics, Arizona State University, Tempe, AZ 85287, USA}
\author{Jesse Kapeghian}
\affiliation{Department of Physics, Arizona State University, Tempe, AZ 85287, USA}
\author{Onur Erten}
\affiliation{Department of Physics, Arizona State University, Tempe, AZ 85287, USA}
\author{Antia S. Botana}
\affiliation{Department of Physics, Arizona State University, Tempe, AZ 85287, USA}

\maketitle

\section{\label{sec:details}Technical details of the DFT  calculations}
Density functional theory (DFT) calculations were performed using the projector augmented wave (PAW) \cite{PAW} pseudopotentials as implemented in the Vienna {\it ab-initio} Simulation Package (VASP) \cite{VASP}.  
In our calculations for the bilayers and monolayers, we have used a $12\times12\times1$ Monkhorst-pack $k$-point grid and plane-wave cutoff energy of 450 eV. 
In our bulk calculations for the rhombohedral phase (see below) we have used an $8 \times 8 \times 4$  Monkhorst-pack $k$-point grid and plane-wave cutoff energy of 450 eV.
When structural relaxations are performed, an additional force convergence criterion of 10 meV/\AA~ was used. 

Different exchange-correlation functionals have been tested as a benchmark for the quality of our calculations (see below). When comparing energies of different magnetic configurations, we account for strong electronic correlation effects for the Cr 3$d$-electrons, using DFT$+U$ within the fully localized version of the double counting correction \cite{Liechtenstein1995}. We set the on-site Coulomb repulsion $U=3$ eV (in agreement with cRPA calculations \cite{mj}) and Hund's rule coupling $J_{\mathrm{H}}$= 0 eV after our benchmarking calculations (see Sec. \ref{sec:functionals}).

\section{\label{sec:bulk}Comparison of Exchange-Correlation functionals for Structural parameters}
We begin by performing geometric relaxations  with different functionals on bulk CrX$_{3}$ compounds to confirm which methodology correctly reproduces the experimental structural parameters. We focus on the rhombohedral phase and perform our relaxations with ferromagnetic (in- and out-of-plane) coupling. We report all structural data below in Table \ref{tab:bulk_calc} and compare them to experiments  from Refs. \onlinecite{McGuire2015, Handy1952, Morosin1964}. We find that both the PBEsol \cite{PBEsol} and the DFT-D3 \cite{DFTD3} van der Waals correction scheme produce structures that are comparable to the experimental data. 

After benchmarking the exchange-correlation functionals for the bulk systems, we turn to the CrX$_{3}$ bilayers. As mentioned in the main text, we construct the initial bilayer structures using a rhombohedral stacking. We perform geometric relaxations for the bilayer structures using both the PBEsol and DFT-D3 schemes. We find that both methods reproduce the experimental data. We have tabulated the lattice parameters from our calcultions and experimental data in Table \ref{tab:latt_params}. We proceed with the PBEsol functional for our total energy calculations in different magnetic states (to obtain the magnetic parameters of Eqs. 2 and 3 in the main text) because the DFT-D3 scheme computes a vdW correction (independent of the spin configuration) that is simply added to the total energy. As such, the vdW energy correction is suppressed when taking energy differences which are necessary for the moir\'e fields and magnetic parameters.

\begin{table*}
\centering
\begin{tabular*}{\columnwidth}{l@{\extracolsep{\fill}}lcccc}
\hline
\hline
Compound (Structure) & Structural parameter & Exp. (Refs. \onlinecite{McGuire2015, Handy1952, Morosin1964}) & GGA & PBEsol & DFT-D3\\
\hline
& $a$ (\AA) & 6.87 & 6.99 & 6.82 &  6.89\\
CrI$_{3}$ ($R\bar{3}$) & $c$ (\AA) & 18.81 & 22.14 & 19.83 & 19.91\\
& $d_{\text{inter}}$ (\AA) & 6.60 & 7.37 & 6.59 & 6.62\\

\hline
& $a$ (\AA) & 6.26 & 6.43 & 6.29 &  6.36\\
CrBr$_{3}$ ($R\bar{3}$) & $c$ (\AA) & 18.20 & 20.57 & 18.7 & 18.30\\
& $d_{\text{inter}}$ (\AA) & 6.10 & 6.85 & 6.23 & 6.09\\
\hline
& $a$ (\AA) & 5.94 & 6.051 & 5.93 &  5.94 \\
CrCl$_{3}$ ($R\bar{3}$) & $c$ (\AA) & 17.33 & 19.44 & 18.01 & 17.23\\
& $d_{\text{inter}}$ (\AA) & 5.78 & 6.48 & 6.00 & 5.74\\

\hline
\hline
\end{tabular*}
\caption{Summary of experimental and calculated structural data from bulk calculations on Cr$X_{3}$ compounds in the rhombohedrally-stacked phase. We have compared three different functionals: GGA, GGA-PBEsol, and GGA-vdW (DFT-D3) to see which one most accurately reproduces the experimental structure. Note that $d_{\text{inter}}$ refers to the interlayer spacing between the Cr honeycomb layers.} \label{tab:bulk_calc}
\end{table*}

\begin{table}
    \centering
    \begin{tabular*}{\columnwidth}{l@{\extracolsep{\fill}}lccc}
    \hline 
    \hline
         Compound  & Parameter & PBEsol & DFT-D3 & Exp.\\
    \hline
         CrI$_{3}$ & $a$ (\AA)           &   6.82   &    6.89   &  6.87 \\
                    & $d_{\text{inter}}$ (\AA) &   6.59   &    6.62   &   6.60 \\
    \hline
        CrBr$_{3}$ & $a$ (\AA)          &  6.29    &  6.36 &   6.26\\
                & $d_{\text{inter}}$ (\AA)&  6.23    &  6.09 &   6.10\\
    \hline
         CrCl$_{3}$ & $a$ (\AA)          &  5.93    &  5.94 &  5.94 \\
                    & $d_{\text{inter}}$ (\AA) &  6.00    &  5.74 &  5.78\\
 
    \hline 
    \hline
    \end{tabular*}
    \caption{Lattice parameters and interlayer distance for CrX$_{3}$ bilayer compounds calculated from first-principles (PBEsol and DFT-D3) compared with the experimental bulk data \cite{McGuire2015, Handy1952, Morosin1964}.}
    \label{tab:latt_params}
\end{table}

\section{\label{sec:functionals}Comparison of exchange-correlation functionals/parameters for magnetic properties}

As mentioned above, when comparing the energies of different magnetic configurations we use DFT+$U$ to account for the strong correlations of the Cr-$3d$ electrons ($U$= 3 eV, consistent with cRPA calculations \cite{mj}). This is the most basic method to deal with correlations, where to the DFT functional is added an orbital-dependent interaction term characterized by an energy scale $U$, the screened Coulomb interaction between the correlated orbitals. DFT+$U$ contains a static description of the electronic interaction neglecting its frequency dependence.
We have systematically explored the functional (PBEsol, optB86-vdW \cite{optB86b}, optPBE-vdW\cite{optPBE}), and Hund's rule coupling ($J_{\mathrm{H}}$) dependence of our DFT+$U$ calculations in different magnetic states (FM planes coupled FM out-of-plane and FM planes coupled AFM out-of-plane) for the CrX$_{3}$ family of materials. Calculations on these two magnetic states were performed in two distinct symmetry inequivalent stackings: AB (rhombohedral) and AB' (monoclinic). Our results are shown in Table \ref{tab:functional}. We find that the magnetic groundstate -and as a consequence the sign and magnitude of the interlayer exchange- predicted from DFT calculations is dependent upon the choice of functional and Hund's coupling $J$. This sensitivity arises from the small energy differences between the FM and AFM spin configurations. As mentioned above and in the main text, we choose PBEsol$+U$ ($U$= 3 eV, J$_H$= 0 eV) since (i) it gives results in bilayers (particularly of CrI$_3$) that agree with experiments (an AFM interlayer coupling is obtained in AB' stacking and a FM one in AB stacking), (ii) it can be used in monolayer calculations for the extraction of $J$, $\lambda$, and $A_s$ (this would not be the case for optPBE-vdW and optB86b-vdW functionals). 
The derived Cr moments are the expected ones $\sim$ 3$\mu_B$ -- the halide ions can also develop a sizable moment, larger for I ($\sim$ 0.13$ \mu_B$), with opposite sign. In an ionic picture, the oxidation state of Cr in these systems is +3, with a 3$d^3$ electronic configuration. The Cr atoms have an octahedral coordination of X ions, so the Cr ions in this environment can be expected to have $S = 3/2$ ($L=0$) with 3 electrons occupying the t$_{2g}$ manifold, as we find. The Cr moment is also consistent with the experimentally observed saturation magnetization \cite{bulkcri3}.

It is relevant to point out that a DFT+$U$ methodology analog to ours established that the stacking order defines the magnetic ground state in CrI$_3$ bilayers \cite{Sivadas2018} and enabled an accurate description on the origin of magnetic anisotropies in monolayers of this material \cite{Lado2017}. These calculations show that  monolayer CrI$_3$  is a ferromagnet with out-of-plane anisotropy, consistent with experiments \cite{Huang2017}.

\begin{table*}[h!]
\begin{tabular*}{\columnwidth}{l@{\extracolsep{\fill}}cccccc}
\hline
\hline
                                                 &   \multicolumn{2}{c}{CrI$_{3}$}        &  \multicolumn{2}{c}{CrBr$_{3}$}       &   \multicolumn{2}{c}{CrCl$_{3}$}       \\   
\hline                                                                                                                                                                     
Functional                                       & MM ($\mu_{B}$)      & $\Delta E$ (meV)      & MM ($\mu_{B}$) & $\Delta E$ (meV)     & MM ($\mu_{B}$) &  $\Delta E$ (meV)       \\
\hline                                                                                                                                                                     
PBEsol$+U$ ($U=3.0$ eV, $J_H=0.5$ eV)            & 3.3   & {-8.61} ({-0.06})  &  3.1  & {-3.51} ({-0.56})& 3.0 & {-1.15} ({-0.11})  \\
optPBE-vdW$+U$ ($U=3.0$ eV, $J_H=0.5$ eV)        & 3.2   & {-6.66} {(0.13)}    & 3.1    & {-2.49} ({-0.21}) & 2.9  & {-0.64} {(0.13)}    \\
optB86b-vdW$+U$ ($U=3.0$ eV, $J_H=0.5$ eV)       & 3.2   & {-7.42} {(0.29)}    & 3.1     & {-5.75} ({-2.97}) & 2.9   & {-0.88} {(0.03)}    \\
$^{\star}$PBEsol$+U$ ($U=3.0$ eV, $J_H=0.0$ eV)  & 3.4   & {-11.77} {(0.61)}   & 3.2     & {-5.16} ({-0.71}) & 3.1  & {-1.95} ({-0.33})  \\
optPBE-vdW$+U$ ($U=3.0$ eV, $J_H=0.0$ eV)        & 3.4   & {-9.79} {(0.84)}    & 3.2     & {-4.12} ({-0.39}) & 3.1   & {-1.41} ({-0.12})  \\
optB86b-vdW$+U$ ($U=3.0$ eV, $J_H=0.0$ eV)       & 3.4   & {-10.5} {(1.05)}    & 3.2     & {-4.59} ({-0.45}) & 3.1 & {-1.67} ({-0.19})  \\
\hline
\hline
\end{tabular*}
\caption{Energy differences ($\Delta E = E_{\text{FM}} - E_{\text{AFM}}$) and Cr magnetic moments for the the CrX$_{3}$  (X = I, Br, Cl) materials at two symmetry inequivalent stackings AB and AB' for several exchange functionals and combinations of Hubbard $U$ and Hund's coupling. The table is organized as $\Delta E^{\text{AB}}$ ($\Delta E^{\text{AB'}}$) in units of meV. A negative energy difference  corresponds to a ferromagnetic exchange coupling, whereas a positive energy difference corresponds to an antiferromagnetic exchange coupling. }
\label{tab:functional}
\end{table*}

\section{Further details on the Interlayer coupling}

\begin{figure}[H]
\centering
\includegraphics[width=0.8\columnwidth]{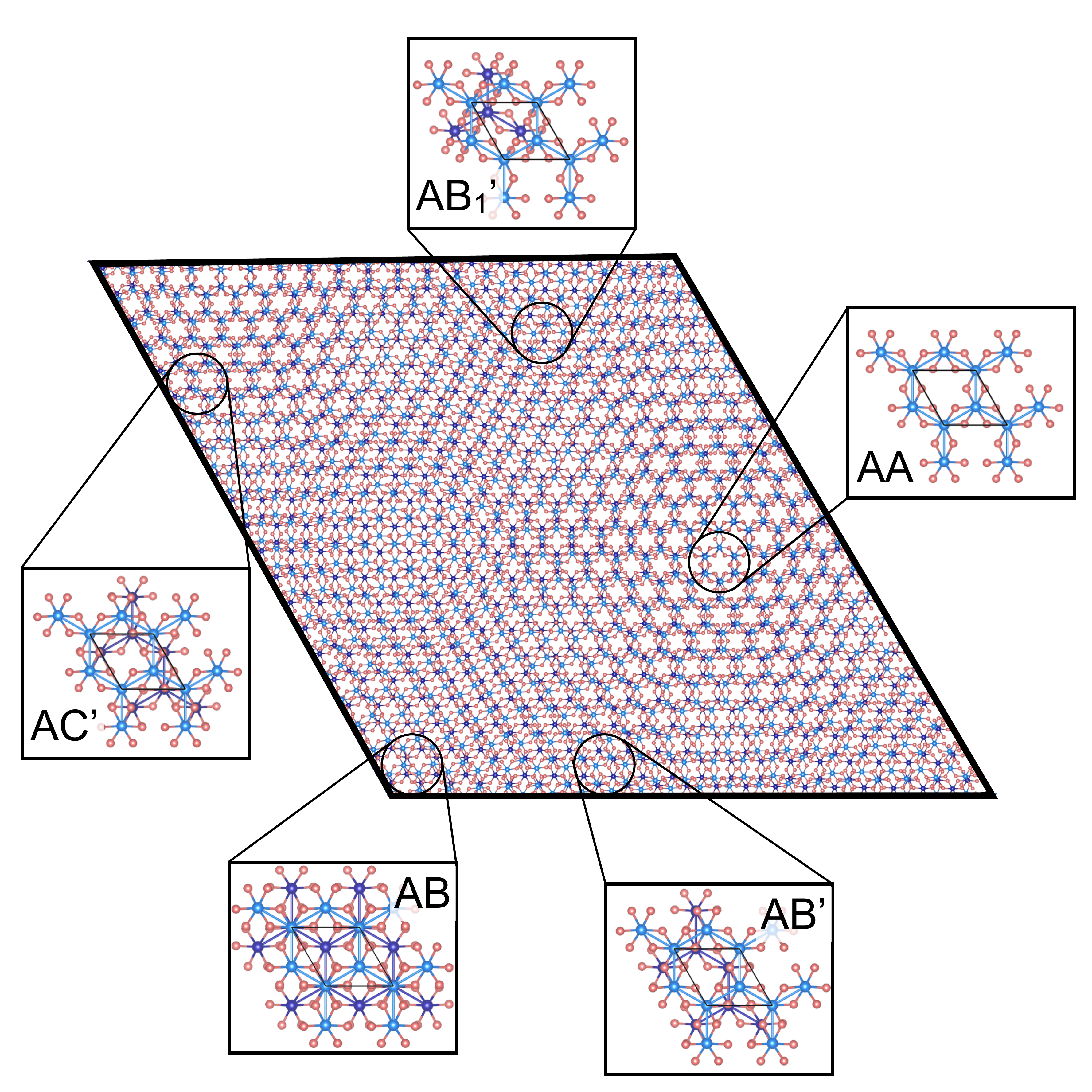}
\caption{Moir\'e pattern of a CrX$_3$ bilayer with a twisting angle of 3$^\circ$. The rhombus represents a moiré unit cell. Local regions resembling lattice commensurate bilayers are named as  AB and AA (symmetric stacking sequences (rhombohedral)) and AB', AB$_{1}$', and AC' (asymmetric stacking sequences (monoclinic)). The larger light (dark) blue spheres represent the Cr atoms in the top (bottom) layer while the smaller, red spheres represent the halide anions.}
\label{fig:moire}
\end{figure}

As mentioned in the main text, for CrX$_3$ bilayers, a small twisting creates a long-period moiré pattern, as schematically shown in Fig. \ref{fig:moire}. The moiré periodicity is approximately
$L$ $\sim$ $a$/$\sqrt{\delta_M^2+\theta^2}$ for small lattice mismatch $\delta_M$ and/or twisting
angle $\theta$, $a$ is the lattice constant of the monolayer.

The stacking-dependent interlayer exchange (J$_\perp$) for our bilayers in the continuum limit calculations is obtained on a 6 $\times$ 6 grid, where each grid point corresponds to a lateral shift ($\frac{i}{6}$, $\frac{j}{6}$) for $i, j = 0$ to 6 along the crystalline axes (as shown in Fig. 1 in the main text). The values on the perimeter are the same due to periodic boundary conditions. We then interpolate this data by fitting to a polynomial in order to get a mesh of $J_\perp$(r). A similar procedure has also been used in Refs. \cite{Sivadas2018, Tong2021}.

\section{\label{sec:dipole}Monolayer calculations and Magnetic anisotropy energies}

As mentioned in the main text, we derive $J$, $\lambda$, and $A_{s}$ from first-principles,
using a procedure analogous to that employed in Ref. \citenum{Lado2017} for monolayer CrI$_3$ considering four magnetic configurations in CrX$_{3}$ monolayers: FM and AFM both oriented in-plane ($x$) and out-of-plane ($z$). These calculations are performed using PBEsol+$U$ ($U$= 3 eV, J$_H$= 0 eV) including spin-orbit coupling (SOC).  The corresponding energies for a unit cell (two Cr atoms) are:
\begin{align}
E_{\text{FM}, z} &= E_{0} -  \Big ( 3(J + \lambda) + 2 A_{s} \Big )|{\bf S} |^{2}\\
E_{\text{AFM}, z} &= E_{0} +  \Big ( 3(J + \lambda) - 2 A_{s} \Big )|{\bf S} |^{2}\\
E_{\text{FM}, x} &= E_{0} - 3J|{\bf S} |^{2}\\
E_{\text{AFM}, x} &= E_{0} + 3J|{\bf S} |^{2}
\end{align}
where $|\vb{S}| = 3/2$ and $E_{0}$ is the reference energy.

The derived values of $J$, $\lambda$, and $A_s$  are shown in Table I in the main text. We focus here on the anisotropy term, given that the magnetic anisotropies play a crucial role in stabilizing various chiral states. 
In general, the anisotropy is described by the magnetic anisotropy energy (MAE), which measures the dependence of the energy on the orientation of the magnetization. There are two origins for the MAE: the magneto-crystalline anisotropy (MCA) that is mainly determined by the SOC, and the shape anisotropy (MSA) ascribed to the classical magnetic dipole-dipole interactions, so that MAE = E$_{\text{MCA}}$+ E$_{\text{MSA}}$. Usually, the MAE can be characterized by the MCA, however, when SOC interactions are weak (like in Cl-based systems) other interactions become relevant and should also be derived. First, we calculate the MCA energy as $E_{\text{MCA}} = E^{\text{SOC, $x$}}_{\mathrm{per ~Cr}} - E^{\text{SOC, $z$}}_{\mathrm{per~ Cr}} \sim (3\lambda/2+A_s)|{\bf S} |^{2}$ with $|{\bf S} |= 3/2$. For all three materials, DFT calculations  predict a positive MCA (corresponding to an out-of-plane easy axis), as  shown in Table \ref{tab:coefficients}. This is in disagreement with experiments for CrCl$_3$ that show an in-plane anisotropy, hinting to the importance of the MSA in this system.  Following Refs. \citenum{Lu2020} and  \citenum{Xue2019}, we calculate the MSA using $E_{\text{MSA}} = E^{\text{dipole}}_{x} - E^{\text{dipole}}_{z}$ (see Table \ref{tab:coefficients}). 
With the inclusion of the MSA, the MAE now shows the anisotropy to be out-of-plane for CrI$_{3}$ and CrBr$_{3}$ and in-plane for CrCl$_{3}$, in agreement with experiments \cite{McGuire2017}. In order to account for this effect, in our effective spin Hamiltonian for CrCl$_3$ we use $A_s/J = -0.007$, correctly accounting for its in-plane anisotropy.

\begin{table}[h!]
\centering
\begin{tabular}{lccc}
\hline
\hline
                             & CrI$_3$  &  CrBr$_{3}$ &  CrCl$_{3}$ \\
\hline

$E_{\text{MSA}}$ (meV/Cr)        & -0.04 & -0.05 & -0.06 \\ 
$E_{\text{MCA}}$ (meV/Cr)        & 0.54 & 0.22  &  0.025 \\
MAE (meV/Cr)                     & 0.50  & 0.17  & -0.035 \\
\hline
\hline
\end{tabular}
\caption{Magnetic anisotropy energies as extracted from DFT calculations. Positive MCA/ MAE values favor an off-plane easy axis. }
\label{tab:coefficients}
\end{table}

More details on the calculations of the magnetic dipole-dipole energy ($E^{\text{dipole}}$) are provided below. It is calculated as given in classical electrodynamics,
\begin{equation}
E^{\text{dipole}} = \frac{1}{2} \frac{ \mu_{0}}{4 \pi} \sum_{i \neq j} \frac{1}{r_{ij}^{3}} \Big ( \vb{m}_{i} \cdot \vb{m}_{j} - 3 \frac{(\vb{m}_{i} \cdot \vb{r}_{ij})(\vb{m}_{j} \cdot \vb{r}_{ij})}{r_{ij}^{2}} \Big ),
\end{equation}
where $\vb{r}_{ij} = \vb{r}_{i} - \vb{r}_{j}$, $\vb{m}_{i}$ is the magnetic moment at site $i$, and $\mu_{0}$ is the magnetic permeability of free space. For a honeycomb lattice with FM moments, the difference between moments in-plane and out-of-plane is
\begin{equation}
E^{\text{dipole}}_{x} - E^{\text{dipole}}_{z} = -\frac{3}{2} \frac{\mu_{0} |\vb{m}|^{2}}{4 \pi} \sum_{i \neq j} \frac{1}{r_{ij}^{3}} \Big ( \frac{ \vb{r}_{ij} \cdot \vu{m}_{i} }{r_{ij}}\Big )^{2}.
\end{equation}
We calculate this energy difference as a function of the size of the lattice until we reach convergence. The results from our calculations the CrX$_{3}$ compounds are shown in Fig. \myblue{2}. Our calculations agree well with previous reported values in Refs. \onlinecite{Xue2019, Lu2020}. 

\begin{figure}[H]
\centering
\includegraphics[width=0.5\columnwidth]{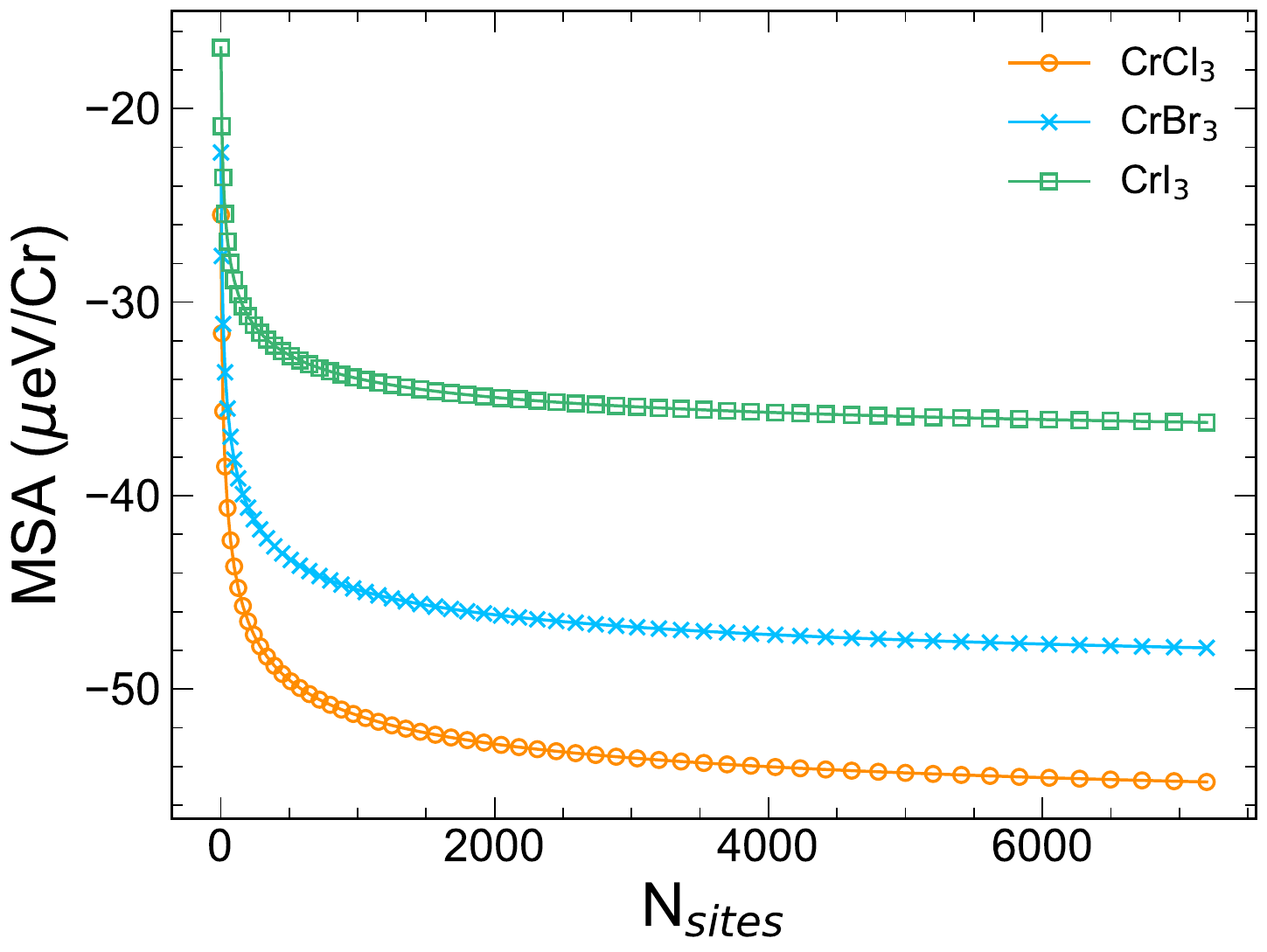}
\label{figs6666}
\caption{Convergence of the MSA energy with respect to the size of the lattice.}
\end{figure}

\section{Continuum free energy}
In order to perform large scale numerical simulations, we take the continuum limit of our bilayer Hamiltonian. The free energy functional is $\mathcal{F}[{\bf m}] = \int d^2r F({\bf m})$, where ${\bf m}({\bf r})$ is the local magnetization. The free energy density $F({\bf m})$ has the following components
\begin{eqnarray}
F({\bf m}) =F_{intra}^1+F_{intra}^2+F_{inter}
\end{eqnarray}
where $F_{intra}^1$, $F_{intra}^2$ are the intralayer free energy densities of layer 1, 2 and $F_{inter}$ is the interlayer exchange free energy density. The intralayer free energy density has the following three components
\begin{eqnarray}
F_{intra} =F_{iso}+F_{DM}+F_{aniso},
\end{eqnarray}
where
\begin{eqnarray}
F_{iso}&=&F_{0}(\textbf{m})+\frac{3}{2}(J/2)\sum_{{\alpha}}(\nabla m^{\alpha})^{2}
\end{eqnarray}
\begin{eqnarray}
F_{DM}&=&-\frac{3}{2}D(m^{z}\partial_{x}m^{x}-m^{x}\partial_{x}m^{z})\nonumber\\
&& +\frac{3}{2}D(m^{y}\partial_{y}m^{z}-m^{z}\partial_{y}m^{y})
\end{eqnarray}
\begin{eqnarray}
F_{aniso}&=&-\frac{3\lambda}{2}(m^{z})^{2}+\frac{3}{2}(\lambda/2)(\nabla m^{z})^{2}\nonumber\\
&&-A_{s}(m^{z})^{2}.
\end{eqnarray}
Here, the isotropic term $F_{iso}$ arises from the isotropic intralayer exchange coupling. The constant term $F_0(\textbf{m})$ in $F_{iso}$ determines the magnitude of magnetization and the second term is the gradient energy whose stiffness $J$ is determined by the intralayer exchange coupling in the discrete model. The second term $F_{DM}$ is the free energy density due to Dzyaloshinskii-Moriya interaction, and the last term $F_{aniso}$ is the free energy density due to the anisotropic exchange coupling and shape anisotropy.
\begin{eqnarray}
F_{inter}&=&-J^{\perp}({\bf r})\textbf{m}^1({\bf r}).\textbf{m}^2({\bf r}).
\label{eqF}
\end{eqnarray}
where $\textbf{m}^{1 (2)}({\bf r})$ is the local magnetization in layer 1 (2) and {\bf r} is the interlayer displacement vector.

We use LLG equations to find the ground state configuration. 
We consider different initial magnetic states pointing along \{(1,1,0), (1,1,.1)\}, \{(1,1,0), (1,1,.2)\}, \{(1,2,0), (2,1,0)\}, \{(1,3,0), (3,1,0)\}, \{(1,1,1), (1,1,-1)\}, \{(1,1,.1), (1,1,-.1)\}, \{(1,1,.1), (1,1,.1)\}, \{(1,1,.01), (1,1,.01)\} and time-evolve the system until the magnetic state is no longer time dependent. We pick the lowest-energy configuration among the different initial states as the ground state configuration. Our calculations are performed on a 1x1 moir\'e unit cell with periodic boundary conditions. The size of the moire unit cell ($L$), is set by the twist angle $L=a/(2\sin(\theta/2))$. The continuum model is solved on a $M\times M$ grid of size $M=24$ and 34 for large angle (small $L$) and $M$=40, 44, 50 for small angle (large $L$). We checked that the convergence is reached with respect to system size. The relative sign or $D$ does not change our phase diagrams. However, the sign of $D$ changes the chirality of the domain walls and the helicity (but not the topological charge) of the skyrmions. Even though the suspended systems have center of inversion in between the layers, there is no center of inversion on the layers. This leads to equal and opposite sign of $D$ for the  suspended systems.

\section{Topological Charge density of $\mathrm{CrCl}_3$}
\begin{figure}[H]
\centering
\includegraphics[width=1\columnwidth]{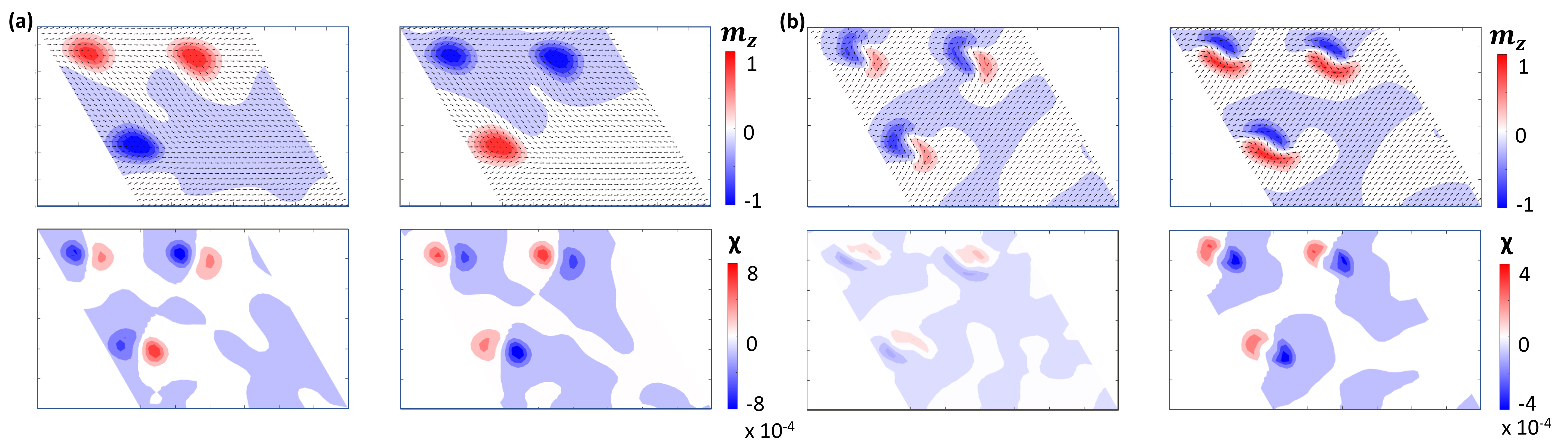}
\caption{(a) Magnetization texture of Hedgehog-1 and its corresponding topological charge density ($\chi$) at $D/J=0.1$ and $\theta=0.48^\circ$. (b) Magnetization texture of Hedgehog-2 and its corresponding topological charge density at $D/J=0.1$ and $\theta=0.2^\circ$.}
\label{fig:top_charge}
\end{figure}

Fig. \ref{fig:top_charge} represents the magnetization textures of Hedgehog-1, Hedgehog-2 and their corresponding topological charge densities. In Hedgehog-1, there is a meron-like spin structure on each AFM patch but the magnetization does not wind properly at the boundary and therefore it does not lead to a quantized topological charge. The topological charge density has a pair of opposite charge patches on each AFM region for $D/J=0.1$ and the net charge is zero. Moreover, the net positive/negative charge increases as we decrease the angle. For large $D/J$ and small $\theta$, the magnetization patterns get slightly deformed. However, the net topological charge in each layer still vanishes. In Hedgehog-2 there are a pair of meron-like spin textures on each AFM patch and similar to Hedgehog-1 the magnetization does not wind properly at the boundaries. The topological charge density also has a pair of opposite charge density patches on each AFM patch and the net topological charge is zero.   

\section{DOMAIN WALL ENERGY AND CRITICAL ANGLE ESTIMATE}
Here, we provide the details of our analytical estimates of domain wall energy and the critical angle. Our results slightly differ from Ref. \citenum{Xu_arXiv2021} which takes a domain wall configuration that completely resides inside the AF region. We take half of the domain wall to reside in the FM region and the other half inside the AF patch.
We also include the contribution from the DM interaction. Following a similar analysis to that in \cite{Xu_arXiv2021}, other than for these distinctions, the domain wall energy per moire unit cell is given by
\begin{equation}
E_{DW} \sim \frac{\pi^2L}{2\delta}\left(J+\frac{\lambda}{2}\right)+\frac{\delta L}{2a^2}\left(A_s+\lambda\right)+\frac{\delta L}{a^2}\left(1-\frac{2}{\pi}\right)\bar{J}^{\perp}-\frac{D\pi L}{a},
\end{equation}
where $\delta$ is the domain wall width. Here, first and second terms are the energy costs in FM region due to the intralayer exchange interactions and magnetic anisotropy, respectively. The third term is the energy cost due to the interlayer exchange interaction. 
Minimizing the domain wall energy with respect to $\delta$ we get the optimal value of the domain wall length,
\begin{equation}
\delta \sim \pi \sqrt{\frac{\left(J+\frac{\lambda}{2}\right)}{A_s+\lambda+0.72\bar{J}^{\perp}}}a.
\end{equation}
Inserting the optimal value of $\delta$ back in $E_{DW}$, we get
\begin{equation}
E_{DW} \sim \frac{\pi L}{a} \sqrt{\left(J+\frac{\lambda}{2}\right)\left(A_s+\lambda+0.72\bar{J}^{\perp}\right)}-\frac{D\pi L}{a}.
\end{equation}
The energy gain (per moir\'e unit cell) of forming an AFM domain wall on AFM regions is 
\begin{equation}
E_{AFM} \sim 2 f_{AFM}\left(\frac{L}{a}\right)^2 \bar{J}^{\perp},
\end{equation}
where $f_{AFM}$ is the area fraction of an AFM patch over the area of the moir\'e unit cell, and $J^{\perp}$ is the average AFM coupling. The critical moir\'e period for flipping an AF domain can be found by equating the interplane exchange energy to the domain wall energy.
\begin{equation}
L_{c} \sim \frac{\pi \left( \sqrt{\left(J+\frac{\lambda}{2}\right)\left(A_s+\lambda+0.72\bar{J}^{\perp}\right)}-D\right)}{2f_{AFM}\bar{J}^{\perp}} a.
\end{equation}
The critical angle is obtained by using $L=a/2\sin({\theta/2}) \sim a/\theta$, and this leads to

\begin{equation}
\theta_{c} \sim \frac{2f_{AFM}\bar{J}^{\perp}}{\pi \left( \sqrt{\left(J+\frac{\lambda}{2}\right)\left(A_s+\lambda+0.72\bar{J}^{\perp}\right)}-D\right)}.
\end{equation}
In our calculations, we use $f_{AFM}=4\%$, $\bar{J}^{\perp}=0.3$meV for CrBr$_3$ and $f_{AFM}=8.33\%$, $\bar{J}^{\perp}=1.3$meV for CrI$_3$. 

\bibliography{ref_resub.bib}